\documentclass[aps,showpacs,prl,twocolumn,amsmath,groupedaddress]{revtex4}

\usepackage{graphicx}
\usepackage{dcolumn}
\usepackage{amsmath}

\newcolumntype{e}[1]{D{.}{.}{#1}}

\newcommand{\state}[3]{$\vphantom{#1}^{#1} \text{#2}_{#3}$}
\newcommand{\units}[1]{\:\mathrm{#1}}                                       
\newcommand{\idx}[1]{_{\mathrm{#1}}}                                        

\DeclareMathOperator{\rot}{rot}
\DeclareMathOperator{\eff}{eff}

\DeclareMathOperator{\PA}{PA}
\DeclareMathOperator{\Rb}{Rb}
\DeclareMathOperator{\Yb}{Yb}

\DeclareMathOperator{\vib}{v}

\begin{document}


\title{Production of ultracold heteronuclear YbRb* molecules by photoassociation}

\author{N. Nemitz}
\author{F. Baumer}
\author{F. M\"{u}nchow}
\author{S. Tassy}
\author{A. G\"{o}rlitz}
\affiliation{Institut f\"{u}r Experimentalphysik, Heinrich-Heine-Universit\"{a}t
 D\"{u}sseldorf, Universit\"{a}tsstra\ss{}e 1, 40225 D\"{u}sseldorf, Germany}


\date{\today}

\begin{abstract}
We have produced ultracold heteronuclear YbRb$^*$ molecules in a combined magneto-optical trap by photoassociation. The formation of electronically excited molecules close to the dissociation limit was observed by trap loss spectroscopy in mixtures of $^{87}$Rb with $^{174}$Yb and $^{176}$Yb. The molecules could be prepared in a series of vibrational levels with resolved rotational structure, allowing for an experimental determination of the long-range potential in the electronically excited state. 

\end{abstract}


\maketitle

  
Ultracold polar molecules offer fascinating prospects for the realization of new forms of quantum matter \cite{Baranov:2002} with possible applications to quantum information \cite{DeMille:2002} and to precision measurements \cite{DeMille:2008,Hudson:2006}. While dense atomic clouds are routinely laser-cooled to $\units{\mu K}$ temperatures, the complex internal structure of molecules has so far prevented the successful application of this direct approach. Among the various approaches currently under investigation \cite{Doyle:2004}, the production of translationally cold molecules from mixed-species ensembles of ultracold atoms is one of the most promising. The possible routes for the conversion from atoms to molecules involve either the use of magnetically tunable Feshbach resonances \cite{Kohler:2006} or light-assisted photoassociation \cite{Jones:2006}. 

While Feshbach resonances allow for an efficient and well-controlled preparation of ultracold heteronuclear molecules in high vibrational levels of the electronic ground state, this method is not applicable to all atomic species. In particular, if one of the atomic species does not possess angular momentum in the ground state, experimentally acccessible Feshbach resonances are typically non-existent \cite{Kohler:2006}. In contrast, production of ultracold molecules by photoassociation is in  principle possible for all combinations of atomic species. 

So far all experimental investigations, which have produced ultracold heteronuclear molecules by photoassociation \cite{Wang:2004a, Mancini:2004a, Sage:2005, Kraft:2006, Schloder:2002}, including the very recent demonstration of optical trapping \cite{Hudson:2008} and of state transfer of Feshbach molecules\cite{Ospelkaus:2008}, have used mixtures of alkalis. In this Letter, we report on the controlled production of ultracold heteronuclear molecules in a mixture of the alkali rubidium (Rb) and the rare earth ytterbium (Yb) in an electronicallly excited state by single-photon photoassociation. While ultimately two or more light fields with different wavelengths will have to be used to produce ultracold ground state molecules \cite{Wang:2004a,Sage:2005}, this is the first decisive step towards the production of a new class of dipolar molecules. The main difference between bialkalis and YbRb is that the ground state of bialkalis is always a $^1\Sigma_0$ while in YbRb it is a $^2\Sigma_{1/2}$ state. This implies that ground state YbRb molecules posess a significant magnetic dipole moment in addition to their electric dipole moment and can thus be trapped and manipulated using magnetic fields. An intriguing prospect for ultracold molecules with an unpaired electron such as YbRb is the realization of lattice spin models \cite{Micheli:2006}.


\begin{figure}[thbp]
\includegraphics[width=8 cm]{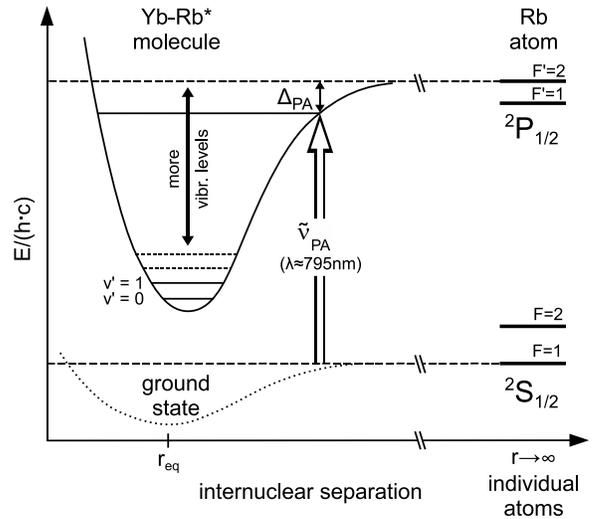}
\caption{
  \label{fig_setup}
Relevant level structure (not to scale) in the Rb atom and the YbRb$^*$ molecule close to the D1-line of Rb at $795\units{nm}$. The detuning from resonance in wavenumbers is defined as $\Delta_{\PA} = \tilde{\nu}_{\PA}-12578.862\units{cm^{-1}}$.}
\end{figure}

\begin{figure*}[thb]
\includegraphics[width=16 cm]{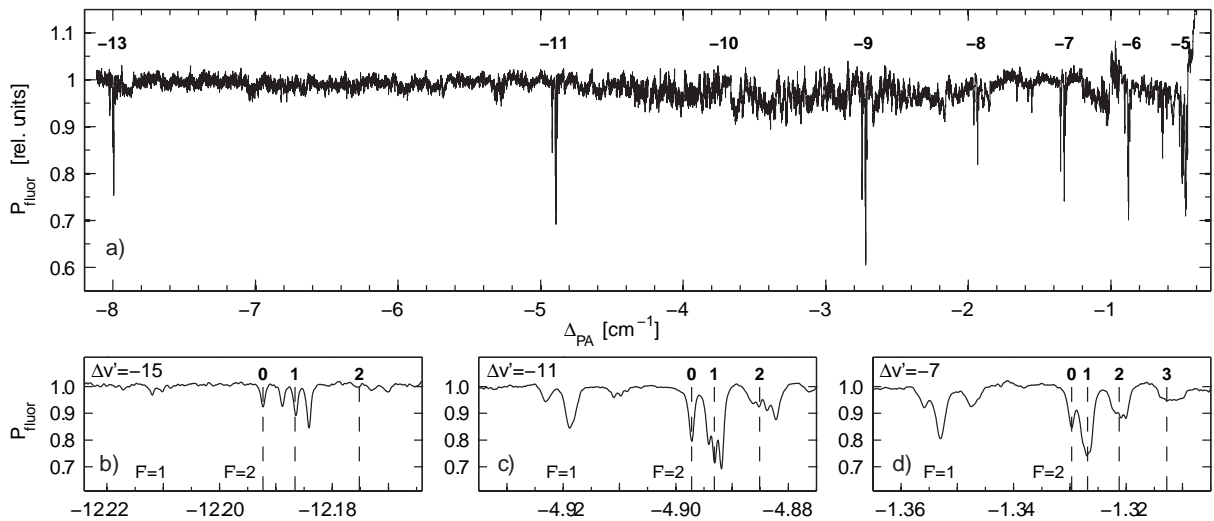}
\caption{
\label{fig_spectra}
a) Partial photoassociation spectrum in a mixture of $^{176}$Yb and $^{87}$Rb. Numbers indicate relative vibrational quantum numbers $\Delta\vib' = \vib'-\vib_{\max}'$. b) - d) Resolved rotational structure for three selected vibrational levels. More tightly bound vibrational levels exhibit a larger rotational splitting and in addition a splitting of the rotational components due to angular momentum coupling.  For all lines two hyperfine components corresponding to the  5$^2$P$\idx{1/2}$, $F'=1$ and $F'=2$ levels of  $^{87}$Rb are observed.}
\end{figure*}

Our experiments were performed using a continuously loaded double-species magneto-optical trap (MOT). Typically, $1.1\times{}10^9$ Rb atoms are trapped in a forced dark-spot MOT \cite{Anderson:1994} which is loaded from a Zeeman slower. The resulting atom cloud has a diameter of $2\units{mm}$ (FWHM) and a temperature of $T_{\Rb} = 340\units{\mu{}K}$, where $>95\units{\%}$ of the atoms are in the dark $F$=1 state. The MOT for Yb operating on the 6\state{1}{S}{0}$\to$ 6\state{3}{P}{1} intercombination transition at $556\units{nm}$ is loaded from a Zeeman-slower operating on the fast 6\state{1}{S}{0}$\to$6\state{1}{P}{1} transition at $399\units{nm}$. It holds $4.3 \times 10^7$ atoms in a $0.5\units{mm}$ cloud at $T_{\Yb} =510\units{\mu{}K}$ when there is no Rb present. With the Rb MOT operating, the number of Yb atoms drops to $4.9\times 10^6$ atoms due to Yb*-Rb collisions. The exponential loading time of the Yb MOT in this situation is typically $0.2\units{s}$.

Here we concentrate on photoassociation close to the 5\state {2}{S}{1/2}$\to$ 5\state{2}{P}{1/2} D1-transition of Rb at $795\units{nm}$. The photoassociation (PA) laser beam is provided by a Ti:Saph laser stabilized to an external resonator that allows for automated scanning over a range of $4\units{GHz}$. Up to $440\units{mW}$ of power are available at the trap position resulting in a peak intensity of $I_{\max}=400\units{W/cm^2}$. Molecules formed by the photoassociation process will generally be lost from the trap \cite{Jones:2006}. Due to the strong imbalance in atom numbers this leaves the Rb MOT virtually unchanged, while the reduction of the number of Yb atoms can be significant. 

In the following, all wavenumbers are given as the detuning $\Delta_{\PA}$ relative to the $F$=1$\to$$F'$=2 transition of the D1 line of Rb as depicted in Fig.~\ref{fig_setup}. Thus, $\Delta_{\PA}$ is a direct measure for the binding energy of the formed molecules. In our measurement sequence, the PA laser is superimposed onto the two overlapped MOTs and swept over its full scanning range with a frequency of $10\units{mHz}$ while the power of the Yb fluorescence $P\idx{fluor}$ is recorded as a measure for the number of trapped Yb atoms. Typically, several sweeps are averaged to obtain a PA spectrum as depicted in Fig.~\ref{fig_spectra}. Since there is no effect of the PA laser on a pure Yb cloud, any decrease of fluorescence can be attributed to Yb-Rb photoassociation \cite{Rb_PA}. 

It is not immediately clear whether the observed photoassociation loss stems from the formation of singly excited YbRb$^*$ or doubly excited Yb$^*$Rb$^*$, since in an Yb MOT, ground and excited state aroms are present. To rule out the formation of Yb$^*$Rb$^*$, we have performed tests in which the atoms were only exposed to the PA light within a periodically recurring dark phase of $50\units{\mu s}$ in which the MOT light is switched off. During this dark phase, only ground state Yb atoms are present. While the efficiency of the MOT is reduced, photoassociation is still clearly observed under this condition, demonstrating that indeed YbRb$^*$ molecules are formed.

Fig.~\ref{fig_spectra} a) shows a partial spectrum for $^{176}$Yb$^{87}$Rb$^*$ constructed from approximately 150 scans. The absolute error of the wavenumber determination is $\pm 5\times 10^{-3}\units{cm^{-1}}$ while the relative position of the components of a vibrational line (Fig.~\ref{fig_spectra} b)\,-\,d)) could be determined with a resolution of $1.6 \times 10^{-4}\units{cm^{-1}}$ which is close to the Doppler broadened linewidth  for an YbRb$^*$ molecule at the effective temperature of $450\units{\mu K}$. 
Only lines for $\Delta_{\PA} < -0.38\units{cm^{-1}}$ could be observed since the PA laser significantly interferes with the Rb MOT performance if its frequency is too close to the atomic resonance. For $\Delta_{\PA} < -8\units{cm^{-1}}$, line positions have been predicted using Leroy-Bernstein methods \cite{LeRoy:1970,Comparat:2004} and only the immediate vicinity has been investigated. We have been unable to find any of the next three lines predicted for $\Delta_{\PA} < -21\units{cm^{-1}}$, probably due to a too small Franck-Condon overlap.  For the strong line at $\Delta_{\PA} \approx -4.9 \units{cm^{-1}}$ in $^{176}$Yb$^{87}$Rb$^*$, we determine a loss rate per Yb atom of $\Gamma_{\PA} = 1.2s^{-1}$  corresponding to a total production rate of excited state YbRb* molecules of $5.9\cdot 10^6\units{s^{-1}}$. This is similar to the results of a comparable experiment with rubidium and cesium \cite{Kerman:2004}, where a trap loss rate of $0.5\units{s^{-1}}$ per cesium atom was measured and it is also in agreement with theoretical predictions based on~\cite{Jones:2006}. 


\begin{table}
\begin{ruledtabular}
\begin{tabular}{e{4}re{2}e{2}e{1}e{4}}
   	\multicolumn{1}{c}{$\Delta_{\PA}$}
&	\multicolumn{1}{c}{$\Delta\vib'$}	
&	\multicolumn{1}{c}{rel.\footnotemark[1]}
&	\multicolumn{1}{c}{$B_{\rot}$}
&	\multicolumn{1}{c}{$r'_{\eff}$} 
&	\multicolumn{1}{c}{ $\Delta_{R'=1}$}					\\
	\multicolumn{1}{c}{ $[\mathrm{cm^{-1}}]$}
&
&	\multicolumn{1}{c}{depth}
&	\multicolumn{1}{c}{ $[\mathrm{cm^{-1}}]$}
&	\multicolumn{1}{c}{$[\mathrm{a_{0}}]$ }	
&	\multicolumn{1}{c}{$[10^{-3} \mathrm{cm^{-1}}]$}							\\
\hline
\multicolumn{6}{l}{\raisebox{1.0ex}{\vphantom{i}}$^{176}$Yb $^{87}$Rb, $F'$=2 state: } \\
- 0.494	& -5	& 0.02	& 0.85	& 34.9	& <0.2\footnotemark[2] 		\\
- 0.881	& -6	& 0.08	& 1.50	& 26.3	& 2.0						\\
- 1.330	& -7	& 0.15	& 1.35	& 27.7	& 0.7						\\
- 1.938	& -8	& 0.09	& 1.70	& 24.7	& 0.4						\\
- 2.723	& -9	& 0.27	& 1.70	& 24.7	& 0.8						\\
- 3.707	&-10	& <0.03	&1.68	& 24.9	& 0.9						\\
- 4.897	&-11	& 0.20	& 2.05	& 22.5	& 1.2						\\
\multicolumn{1}{c}{---}		&-12	& \multicolumn{4}{l}{not observed}		\\
- 8.001	&-13	& 0.16	& 2.50	& 20.3	& 1.6						\\
- 9.950	&-14	& 0.02	& 2.63	& 19.9	& 2.0					 	\\ 
-12.192	&-15	& 0.08	& 2.92	& 18.8	& 2.4						\\
-14.808	&-16	& 0.06	& 3.05	& 18.4	& 2.5					 	\\
-17.687	&-17	& 0.05	& 3.16	& 18.1	& 2.6						\\
-20.921	&-18	& 0.10	& 3.30	& 17.7	& 2.9						\\
\hline
\multicolumn{6}{l}{\raisebox{1.0ex}{\vphantom{i}}$^{174}$Yb $^{87}$Rb, $F'$=2 state: } \\
-0.425	& -4	& 0.15	& 0.85	& 35.0	& <0.5\footnotemark[2]		\\
-0.728	& -5	& 0.10	& 1.15	& 30.1	& <0.5\footnotemark[2]		\\
-1.149	& -6	& 0.18	& 1.40	& 27.2	& <1.0\footnotemark[2]		\\
-2.437	& -8	& 0.17	& 1.67	& 24.9	& 0.68					\\
-4.459	&-10	& 0.15	& 1.95	& 23.1	& 1.08					\\
-7.384	&-12	& 0.11	& 2.33	& 21.1	& 1.65					\\				
\hline
\multicolumn{6}{l}{accuracy: } \\
\pm 0.005& &\pm 0.03 &\pm 0.08& 		& \pm 0.2\\ 
\end{tabular}
\end{ruledtabular}
\footnotetext[1]{relative loss of fluorescence for $R'$=0 component}
\footnotetext[2]{not resolved}
\caption{\label{tab_wn}Properties of observed lines in $^{176}$Yb$^{87}$Rb$^*$ and $^{174}$Yb$^{87}$Rb$^*$. $\Delta\vib'=\vib'-\vib'_{\max}$ is the relative vibrational quantum number, the effective internuclear distance $r'_{\eff}$ is related to the rotational constant $B_{\rot}$. $\Delta_{R'=1}$ is the spacing between adjacent components of the $R'$=1 rotational line.}
\end{table}
  
The majority of observed lines belongs to a vibrational series converging on the excited 5\state{2}{P}{1/2} Rb state. Each vibrational level shows two separate rotational progressions corresponding to the $F'$=1 and $F'$=2 states of the Rb atom. The hyperfine splitting does not significantly vary from the atomic value of $0.0273\units{cm^{-1}}$ over the observed range. Table~\ref{tab_wn} lists the wavenumbers and relative strengths for the stronger $F'$=2 components. The strong fluctuation of the line depth may be caused by oscillations of the Franck-Condon factor. Because of the shorter range of the excited state potential in heteronuclear molecules as compared to homonuclear molecules, these happen on a smaller energy scale comparable to the vibrational spacing~\cite{Azizi:2004}. 

\begin{figure}[t!bh]
\includegraphics[width=8 cm]{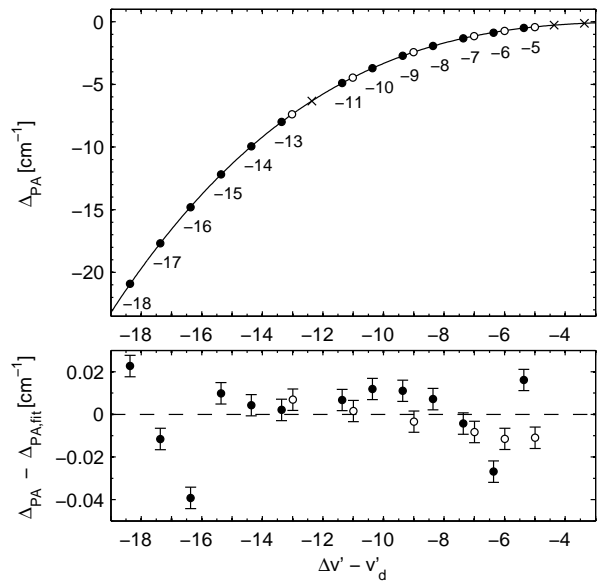}
\caption{\label{fig_leroy}
\textbf{(top)} Energies $\Delta_{\PA}$ of the vibrational $F'$=2 lines over vibrational level relative to non-integer dissociation level ($\Delta\vib'-\vib'_d$). Vibrational levels for $^{174}$Yb ($\circ$) are scaled by $\sqrt{\mu_{176} / \mu_{174}}$ to account for the difference in the reduced mass. Levels for $^{176}$Yb ($\bullet$) are labeled with relative quantum numbers $\Delta\vib'=\vib'-\vib'_{\max}$. Predicted but unobserved levels in $^{176}$Yb are marked  `x'. The solid line is a fit to the data according to the model of Ref.~\cite{Comparat:2004}.  \textbf{(bottom)} Difference between observed values and fit.\\}
\end{figure}

All observed lines show resolved rotational components according to $E_{\rot} = B_{\rot} R' (R'+1)$ where $B_{\rot}$ is the rotational constant and $R'$ is the quantum number for the rotation of the nuclei in the excited state as illustrated  in Fig.~\ref{fig_spectra} b)\,-\,d). Due to the temperature of only several $100\units{\mu K}$ only $R'=0,1,2$ components were observed for most of the lines since the ground state centrifugal barrier for the $R$=3 component at $114\units{a_0}$ has a height of $916\units{\mu K}$. In the fixed rotor approximation, the rotational constant is $B_{\rot} = \hbar^2 / (2\, \mu \,  r)$, where $\mu$ is the reduced mass and $r$ is the distance of the nuclei. For the low rotational quantum numbers in our experiment, centrifugal stretching effects can be neglected and an effective fixed-rotor radius $r'_{\eff}=\hbar / (2\mu{}B_{\rot})$ may be defined for the excited YbRb$^*$ molecules, which is  also listed in table~\ref{tab_wn}.

The finest structure in the line structure is a splitting of the rotational components. The observed pattern agrees with a Hund's case (e) angular momentum coupling, where the total nuclear and electronic angular momentum $F'$ is that of the excited Rb atom ($F'$=1 or $F'$=2) with no contribution from Yb. The angular momentum $F'$ then couples to the nuclear rotation $R'$. As Fig.~\ref{fig_spectra} b) and c) show, the number of observed components is compatible with the expectations for this case. The splitting $\Delta\idx{R'}$ between adjacent components generally increases for lower, more tightly bound vibrational states as expected due to the stronger coupling of the rotation to the electronic angular momentum. 
Taking the whole observed structure into account, the wavenumbers for the individual line components are related to the experimentally determined constants in table~\ref{tab_wn} by
\begin{equation}
\tilde{v}  = \tilde{v}\idx{res} + \Delta_{\PA} + B_{\rot} R' (R'+1) + m'\idx{R'} \Delta\idx{R'}.
\end{equation}
where $m'\idx{R'}$ runs from -$R'$ to $R'$ (or -$F'$ to $F'$ for $R'>F'$).

In the near-dissociation limit, the vibrational energies are predominantly determined by the long-range dispersion coefficients. The improved Leroy-Bernstein method as described in Ref.~\cite{Comparat:2004} has been used to assign vibrational quantum numbers and extract values for the dispersion coefficients. Since the total number of vibrational levels in the potential well is unknown, table~\ref{tab_wn} lists quantum numbers $\Delta\vib'=\vib'-\vib'_{\max}$ relative to the last vibrationalelvel. As Fig.~\ref{fig_leroy} illustrates, the observed line positions are reproduced by a fit to the theoretical model of Ref.~\cite{Comparat:2004}. The fit \cite{LB_fit} yields 
$\vib'_d(^{176}\mathrm{Yb}^{87}\mathrm{Rb}^*)=0.365$ and $\vib'_d(^{174}\mathrm{Yb} ^{87}\mathrm{Rb}^*)=0.991$ for the noninteger values of the relative vibrational quantum numbers at the dissociation limit while the resulting dispersion coefficients are $C'_6 = 6190 \units{a.u.}$ and $C'_8 = 403000 \units{a.u.}$ similar to the values predicted for other heteronuclear atom pairs~\cite{Marinescu:1999}. 

\begin{figure}[tbhp]
\includegraphics[width=8 cm]{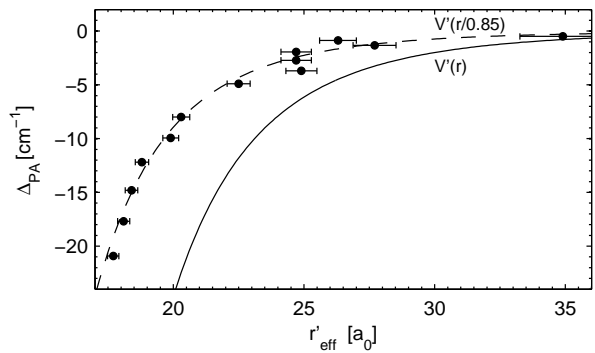}
\caption{\label{fig_reff} Comparison of $r'_{\eff}$ as determined from the rotational constants ($\bullet$) to the potential curve V'(r) given by $C'_6$ and $C'_8$ coefficients (solid line) for $^{176}$Yb$^{87}$Rb$^*$. The dashed line is the potential scaled to 85\% horizontally. 
}
\end{figure}

We have tested our interpretation of the rovibrational structure of the YbRb$^*$ moelcule by comparing the long-range potential $V'(r) = -C'_6/r^6 - C'_8/r^8$ from the Leroy-Bernstein fit to the effective internuclear distances $r'\idx{eff}$ as obtained from the rotational structure. For a given vibrational wavefunction $\psi\idx{v'}(r)$, $r'\idx{eff}$ may be approximated by $r'_{\eff} = (\int_{-\infty}^{\infty} \psi\idx{v'}(r)^2 \, r^2 \, dr)^{0.5}$. While $r'_{\eff}$ is  always smaller than the classical outer turning point $r'\idx{max}$ given by $\Delta\idx{PA} = V'(r'\idx{max})$, the vibrational wavefunction is concentrated near $r'\idx{max}$ for levels close to the dissociation limit and thus $r'_{\eff}$ approaches $r'_{\max}$.  Fig. \ref{fig_reff} demonstrates that our analysis qualitatively agrees with this argument, since the values for $r'\idx{eff}$ fall on a curve corresponding to the Leroy-Bernstein potential scaled by $85\%$. 

In conclusion, we have produced ultracold electronically excited YbRb$^*$ molecules in well defined rovibrational levels by photoassociation. By precise determination of the position of the rovibrational levels close to the dissociation threshold, we were able to model the long-range part of the molecular potential. These results are an invaluable first step towards the production of ultracold ground state YbRb molecules which will involve two-color photoassociation to high-lying vibrational levels in the electronic ground state and subsequent transfer to low-lying vibrational levels as has recently been demonstrated for homonuclear molecules \cite{Danzl:2008}. In the next step, we will combine the photoasociative production of ultracold molecules with conservative trapping of the Yb-Rb mixture as we have recently demonstrated \cite{Tassy:2008}. 

We acknowledge stimulating discussions with T. Fleig. The project is supported from DFG under 
SPP 1116. F.B. was supported by a fellowship from the Stiftung der Deutschen Wirtschaft.


\begin{thebibliography}{27}
\bibitem[{Rb_()}]{Rb_PA}
\bibinfo{note}{Rb-Rb photoassociation only has a small effect on the Yb signal,
  leading to a slight increase of the Yb fluorescence due to reduced Rb-induced
  loss.}
\end{thebibliography}

\end{document}